**Delineating site response from microtremors: A case study**

R. Biswas*, N. Bora

Department of Physics, Tezpur University, Tezpur-784028, Assam

*Corresponding author: rajib@tezu.ernet.in*

**Abstract:**

*We report estimation of site response in the form of fundamental frequency. Towards this objective, we deploy widely established receiver function technique. Taking locally recorded events as inputs, we implement this technique to estimate resonance frequency in three receiver sites, characterized by varying lithology underneath. It is observed that resonance frequencies varies from 3 to 7 Hz, which is also confirmed by our previous studies of estimates from ambient noise recordings with reference to identical sites. Variation of frequency implies existence of heterogeneity in the study area.*

# 1 INTRODUCTION

Estimation of spatial variation of site response is one of the prime objectives of microzonation studies. Several literatures emphasized the paramount significance of site dependent factor which largely impact the concentration of damage in some pocket areas pertaining to local geology (King and Tucker, 1984; Aki, 1988; Kawase, 1998; Field and Jackob, 1993; Nath et al., 2000, 2002a, b). Even the shaking of man-made structures also get influenced by the spatial variation of site effects. The site effects can be parameterized by factors like resonance frequency, amplification factor which have direct impact on semi-resonance or resonance of certain building types, and thereby causing distortions or failures of constructions. In order to compute site response, there are adoptions of various methodologies (Castro et al., 1990; Field and Jacob, 1995; Bonilla et al., 1997; Riepl et al., 1998). HVSR or receiver function technique has been widely regarded as one of the most powerful methods available to seismologist for acquiring a reliable estimate of site response.

In this work, we endeavor to estimate site response through horizontal to vertical ratio of microtremors. We give a comprehensive analysis of the results attained through horizontal to vertical ratio of locally recorded waveforms.

# 2 DATA

In order to generate local waveforms exclusively for this study, a temporary network of three stations namely IIG, NEHU and SETUK were installed in Shillong City. This network was operational for two months. The stations were equipped with three Trillium 120P seismometers from Nanometrics having frequency bandwidth of 0.003 to 50 Hz; with 24 bit Guralp Digitizer in synchronization with Guralp GPS. It was a continuous mode of recording in all the three stations. The data were digitized at a sampling frequency of 100 samples /second. The seismic stations are shown in Figure 1. A total of 135 tremors were recorded during the period of deployment of this temporary network. Out of this, a total of 40 tremors recorded by the three stations were precisely located adopting the velocity model of Bhattacharya et al., (2005), compatible for Shillong region with a view to determine the hypocentral parameters. Out of these 40 events, only fourteen events have been selected in order to study the site response from HVSR in this study region within an epicentral distance of less than 50 km. Table 1 provides the hypo-central parameters of the located events. The depths of the events vary from 4 to 25 km whereas the epicentral distance ranges from a mere 1.9 km to 48 km. The root mean square of the located events is below 0.2.

## 2 HVSR Estimation

As per reports of Aki, 1988; Kawase, 1998, microearthquake study entailing events at shot epicentral distances facilitates the understanding of physics of source processes as well as local site conditions. HVSR is based on the assumption that vertical component is least affected by near-surface influence. Consequently, when we divide the horizontal component by vertical component, the site effect can be deciphered.

### 2.1 Theoretical Background

As an input for HVSR, we incorporate S-wave packets. Suppose, $S(r_{lm}, f_n)$ and $T(r_{lm}, f_n)$ represent S-wave amplitude and the background noise amplitude, respectively with a hypo central distance of $r_{lm}$. The relation between signal amplitude $A_{lm}$ with respect to frequency $f_n$ emerges to be,

$$A(r_{lm}, f_n) = S(r_{lm}, f_n) - T(r_{lm}, f_n) \tag{1}$$

Assuming *k* events being recorded by *j* stations, the amplitude spectrum $A(r_{lm}, f_n)$ in frequency domain of $f_n$ can be expressed as (Lermo et.al., 1993, Nath et al., 2002a; Mandal et al., 2005)

$$A(r_{lm}, f_n) = SI_k(f_n) \cdot P(r_{lm}, f_n) \cdot SO_k(f_n) \tag{2}$$

The corresponding HVSR can be estimated as

$$\text{HVSR}_{kj}(f_n) = \frac{\frac{1}{\sqrt{2}} \sqrt{absH_{kj}(f_n)|^2_{NS} + absH_{kj}(f_n)|^2_{EW}}}{absV_{kj}(f_k)} \tag{3}$$

where $H_{kj}(f_n)|^2_{NS}$, $H_{kj}(f_n)|^2_{EW}$ and $V_{kj}(f_n)$ represent the Fourier spectra of the North-South component, East-West and Vertical component, respectively. By taking into account the contribution of all the seismic events, the average receiver function $\text{HVSR}_j^{ave}(f_n)$ is estimated.

## 2.2 Approach

HVSR yields a peak with existence of an impedance contrast. The epicentral plot of the events used for implementing this receiver function is illustrated in Figure 1. The horizontal to vertical ratio technique adopted for the locally recorded earthquakes in the present study is described below:

i) Differentiation of the S-wave motions from multiple station

ii) Instrument response correction with incorporation of poles and zeroes

iii) Adoption of band pass filter in the range of 0.1-10 Hz.

iv)  the S-wave packets recorded are windowed with a window width containing the maximum amplitude. The window length was selected following the results of Seekings et al., (1996). The resultant multicolumn files containing the corrected displacement spectra are made input in the Matlab code modified after Nath et.al. (2002a). Through the Matlab code, some additional processing are accomplished, including the tapering applied to the time-windowed data.

During the estimation approach of HVSR, we ensure that the corrected spectra is endowed with noise to signal ratio of less than a factor of 3 to eliminate all sorts of plausible transients, as followed by Nath et al., 2002a.

## 3 RESULTS

The receiver function was determined at the three temporary stations viz, SETUK, NEHU and IIG incorporating the waveforms recorded by this temporary network. All these stations are characterized by different type of site geology. The HVSR yields different type of amplification levels and peak frequency corresponding to highest amplification. All these are elaborate station wise.

**STATION: IIG**

The average HVSR result for IIG station including all the local events utilized are outlined between 4 and 6.8 Hz, as evident from Figure 2a. In between 3 and 5 Hz, an appreciable level of amplification is found which later on decays. But, with increment of frequency the amplification rises again and reaches the peak at 1.285. The frequency corresponding to the highest amplification, generally referred to as the fundamental frequency, is hence found to be 6.4 Hz. Meanwhile, the level of amplification remains above one, although very low, for this entire range of frequency. It is worthwhile to note that for higher side of frequencies, no amplification is observed.

**STATION: SETUK**

Similarly, the average HVSR for SETUK station is also estimated which is shown by Figure 2b. The range of frequency outlining the amplification levels is observed to be from 5 to 9 Hz. Interestingly, SETUK reveals a different pattern of site response. Drastic decline of site amplification is contemplated once the peak frequency entailing the highest site response is crossed. SETUK station is characterized by peak frequency of 8Hz. Towards higher side of frequencies above 9 Hz, site amplification is quite negligible. The site response hardly goes beyond 1.0 except at the peak frequency within the frequency band observed in this case. Site amplification is found to be in the range of 0.8 to 1.0 only.

**STATION: NEHU**

Likewise, the average receiver function is also evaluated for NEHU station exploiting the local waveforms recorded during its deployment exclusively for this study. Here also, the average HVSR shows appreciable site response in the same band of frequencies as has been observed for SETUK, i.e; 5 to 9 Hz, as illustrated in Figure 2c. Point of dissimilarity in these two stations arises with the inferred peak frequencies. The average HVSR reveals a fundamental or so called peak frequency at 7.4 Hz. However, the site amplification exhibited by this station is quite higher in comparison to the other two sites. Mostly, the site response is coming within the range of 0.95 to 1.6 which is not usually the case seen for the other two sites. More importantly, the range of frequencies within which the site amplification attains its highest level comes out to be the same for station SETUK and NEHU. It is indicative of the fact that both these sites might be characterized by existence of basement rock strata at the same level of depth.

## 5 DISCUSSION

In this study, emphasis was given to estimate the site response from computation of HVSR. The earthquakes utilized in this study indicate the level of seismic hazard in Shillong City for damage and destruction. The city of Shillong has experienced no. of felt earthquakes in the recent past and has seen huge urbanization during last two decades. This has called an extensive study with the incorporation of various ground parameters like geology, topography, subsoil condition, geomorphology, seismicity, site amplification behavior and the attenuation parameter. Extensive work throughout the world has been carried out by different researchers so as to discern the spatial variability of seismic responses with HVSR formulation-receiver function technique (Langston, 1977; Lermo et al., 1993). It is noteworthy to mention that all of their findings strongly put HVSR as the effective approach while estimating the site response in context of fundamental frequency of receiver site. As depicted in Figure 2, we can observe different type of amplification levels and peak frequencies as yielded by the average HVSR results. This implies frequency dependence of site response with three stations in consideration. While we find lower magnitude of amplification in high frequency range, the opposite trend prevails in low frequency range. It is found that resonance frequency is mostly prominent in the range of 4-8 Hz when all the three receiver sites are taken into account. The IIG site reached its peak at 5.5 Hz as while the rest two sites attained the highest amplification levels attain maximum corresponding to frequencies of 6-7 Hz for two sites. This observation is found to be in good conformity with the estimates of fundamental frequencies estimated through ambient noise recordings at these same sites as reported in Biswas and Baruah, 2011and Biswas et al., 2015a, Biswas and Baruah, 2015b, 2016. As documented in these reports, the horizontal to vertical ratio by modified technique of Nakamura, 1989 & 2008 showed identical estimates of fundamental frequencies for these sites. Thus, it can be inferred that receiver function technique effectively reveals the fundamental frequencies at the sites under investigation. The estimations also implicates that there are spatial variation in geology and soil conditions prevailing in the region.

# 6 CONCLUSION

The site response has been estimated from locally recorded events with adoption of receiver function technique. The fundamental frequencies are found to be in the range of 3 to 7 for the study area pertaining to three receiver sites. The variation in resonance frequencies as computed from this technique implicates prevalence of heterogeneity in the study area. We find good corroboration between estimates of fundamental frequencies from receiver function technique with reference to local events and as that of ambient noise records for the study area. The study will help in future course of mitigation studies in this region.

Figures:

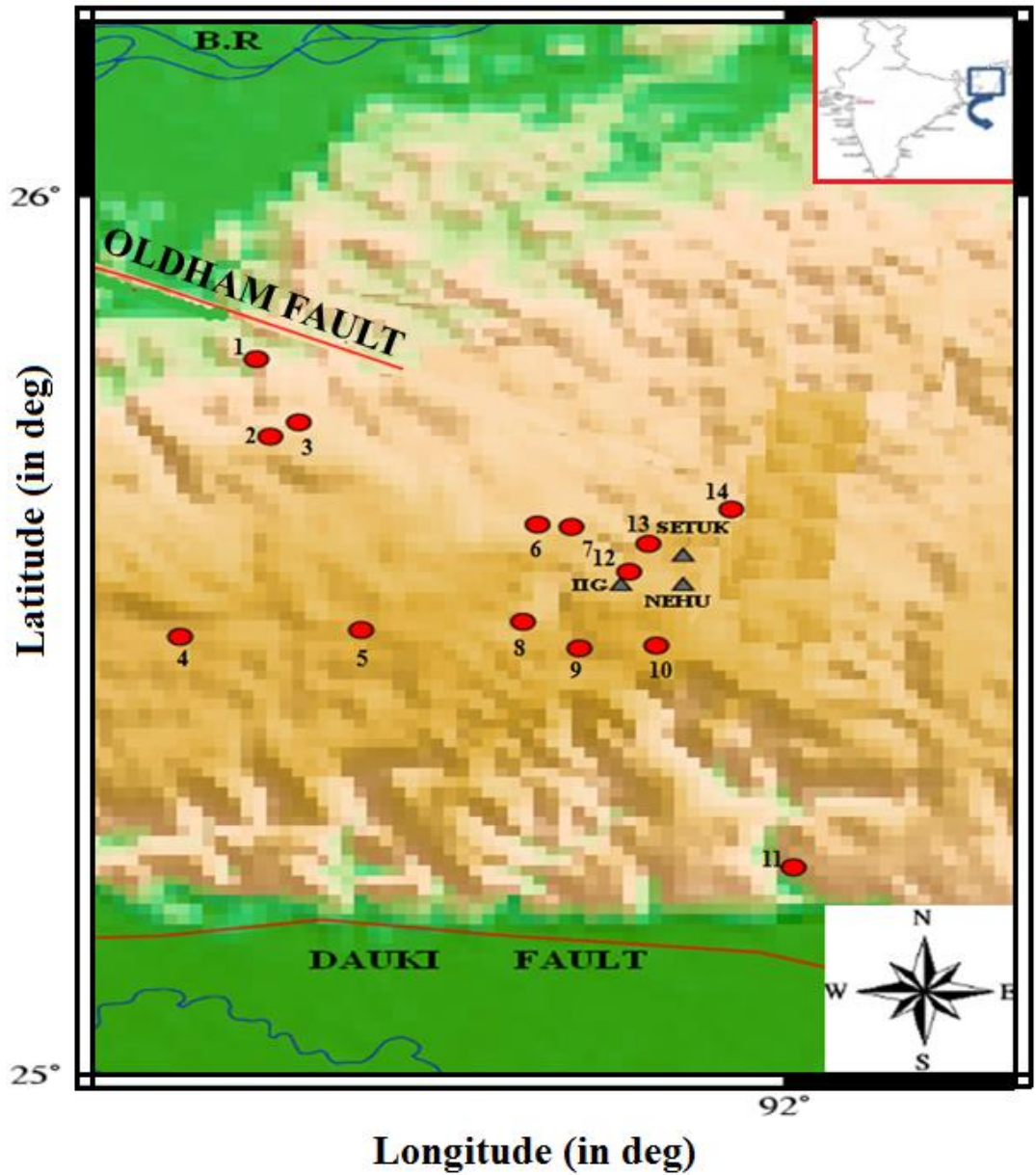

Figure 1 Epicentral plot of the locally recorded events denoted by filled circles. The receiver sites are represented by the filled triangles. BR stands for Brahmaputra River. The study area is shown in the inset.

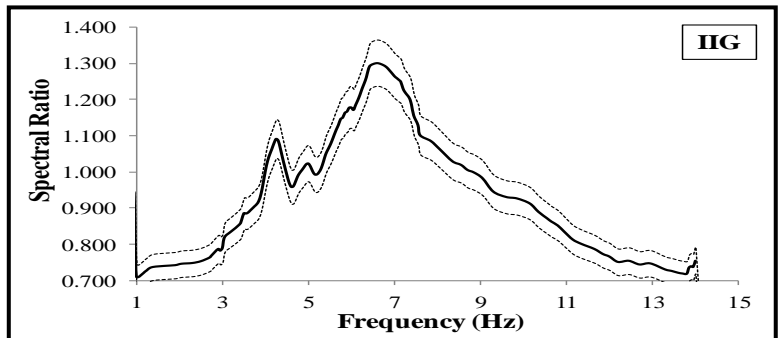

(a)

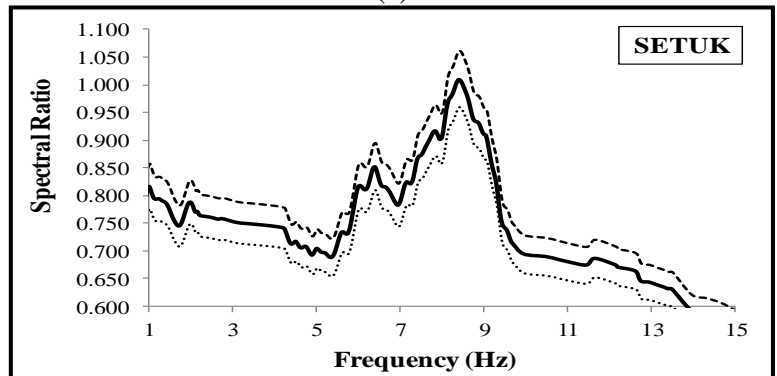

(b)

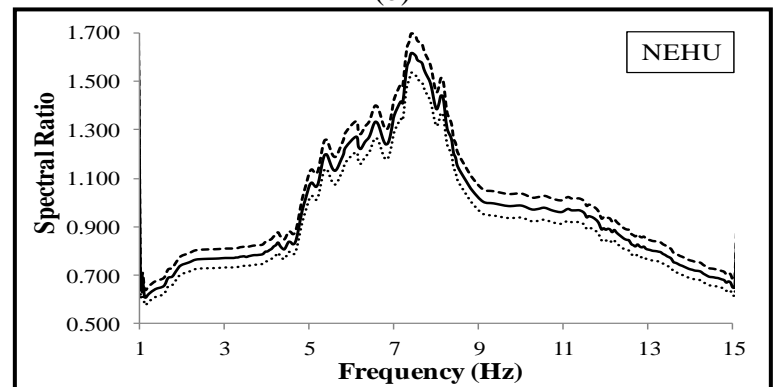

(c)

Figure 2 HVSR estimates at the three temporary stations
    (a) HVSR estimates for station IIG;
    (b) HVSR estimates for station SETUK
    (c) HVSR estimates for station NEHU. The solid line represents the average
    estimate whereas the dotted lines indicate ±5% deviation Table 1

Table 1 List of the events

| Date yymmdd | Origin Time hh mm ss | Latitude in deg | Longitude in deg | Depth km | Mag |
|---|---|---|---|---|---|
| 09 03 09 | 15 33 45.39 | 25.816 | 91.542 | 21.81 | 2.38 |
| 09 03 09 | 18 37 49.45 | 25.744 | 91.579 | 32.87 | 2.26 |
| 09 03 09 | 18 37 49.45 | 25.728 | 91.554 | 30.24 | 2.60 |
| 09 03 19 | 17 45 12.73 | 25.237 | 92.008 | 35.04 | 2.48 |
| 09 03 20 | 21 03 32.47 | 25.574 | 91.865 | 20.00 | 1.85 |
| 09 03 22 | 19 28 57.65 | 25.645 | 91.954 | 09.53 | 1.62 |
| 09 03 27 | 35 02 35.41 | 25.625 | 91.815 | 04.09 | 1.64 |
| 09 03 28 | 09 05 06.80 | 25.500 | 91.476 | 21.67 | 2.51 |
| 09 03 29 | 17 03 08.63 | 25.508 | 91.633 | 20.80 | 2.39 |
| 09 03 30 | 02 31 38.27 | 25.490 | 91.889 | 18.60 | 2.17 |
| 09 03 30 | 07 31 28.26 | 25.628 | 91.786 | 21.00 | 1.44 |
| 09 03 30 | 19 55 05.92 | 25.606 | 91.882 | 13.30 | 1.55 |
| 09 03 31 | 10 15 38.23 | 25.487 | 91.822 | 19.55 | 2.05 |
| 09 03 31 | 21 26 16.86 | 25.517 | 91.773 | 20.80 | 1.87 |